\documentclass[acmconf]{acmart}
\usepackage{listings}
\usepackage{parcolumns}
\usepackage{wrapfig}

\begin{document}

\title{MPI Advance : Open-Source Message Passing Optimizations}

\author{Amanda Bienz}
\authornote{All authors contributed equally to this research.}
\email{bienz@unm.edu}
\orcid{1234-5678-9012}
\affiliation{%
  \institution{Dept.\@ of Computer Science \\
   Univ.\@ of New Mexico}
  \city{Albuquerque}
  \state{NM}
  \country{USA}
}

\author{Derek Schafer}
\authornotemark[1]
\email{dschafer1@unm.edu}
\affiliation{%
  \institution{Center for Advanced Research Computing \\
   Univ.\@ of New Mexico}
  \city{Albuquerque}
  \state{NM}
  \country{USA}
}

\author{Anthony Skjellum}
\authornotemark[1]
\email{askjellum@tntech.edu}
\affiliation{%
  \institution{
    Dept.\@ of Computer Science \\
   Tennessee Technological University} 
  \state{TN}
  \country{USA}
}

\begin{abstract}
The large variety of production implementations of the message passing interface (MPI) each provide unique and varying underlying algorithms.  Each emerging supercomputer supports one or a small number of system MPI installations, tuned for the given architecture.  Performance varies with MPI version, but application programmers are typically unable to achieve optimal performance with local MPI installations and therefore rely on whichever implementation is provided as a system install.  This paper presents MPI Advance, a collection of libraries that sit on top of MPI, optimizing the underlying performance of any existing MPI library.  The libraries provide optimizations for collectives, neighborhood collectives, partitioned communication, and GPU-aware communication.
\end{abstract}


\makeatletter
\let\@authorsaddresses\@empty
\makeatother

\maketitle

\section{Introduction}
A large variety of message-passing interface (MPI) libraries are currently in production, including OpenMPI, MPICH, MVAPICH, and various proprietary implementations.  Each emerging supercomputer provides a system install of one or a small number of MPI implementations, tuned to obtain optimal performance for a given architecture.  While each MPI implementation provides the standard API, underlying implementations vary drastically.  As a result, the performance of parallel applications is dependent on not only the architecture on which they are run but also the implementations within the available system MPI install.  While additional versions of MPI can be installed through package managers, such as Spack, performance will typically be subpar in comparison to tuned system installations.  In this paper, we introduce MPI Advance\footnote{\url{https://github.com/mpi-advance}}, a collection of lightweight libraries that sit on top of MPI, providing advanced algorithms and new MPI features while also leveraging the tuned performance of the system MPI.%

There are many benefits to lightweight libraries that sit on top of MPI, such as MPI Advance. The first is that the simplicity of this approach allows users to experiment with research MPI extensions without having to edit production MPI releases, allowing libraries to be portably tested across various MPI implementations and computer architectures more directly. Such a design also allows for the creation of libraries that applications can utilize to use communication optimizations, such as emerging algorithms for collective operations and locality-aware aggregation~\cite{BienzLocAwareSpMV2019,BienzLocAwareAMG2020,LockhartLocAwareECG}. MPI Advance also provides the opportunity for users to add new MPI methods or optimizations that are not yet available within the MPI standard, allowing for testing within a variety of applications to gather evidence for whether new ideas should be added to the standard. Finally, the libraries allow for implementations of new additions to the standard, such as partitioned communication~\cite{MPIPCL-JOURNAL}, to be made available to application users before system MPI installs have been updated. The end goal of most MPI Advance libraries will be that the learning associated with creating the library can make the transition of the library into a production MPI release more streamlined.

The remainder of this paper describes MPI Advance in detail. Section~\ref{section:impl} outlines what libraries are currently included within MPI Advance, along with the specific rationale for being included. Benefits and preliminary performance results of these libraries are then described in Section~\ref{section:results}. Finally, Section~\ref{section:conc} provides concluding remarks.

\section{MPI Advance}~\label{section:impl}
MPI Advance provides a framework for distributing communication optimizations, including optimizations to collective algorithms and neighborhood collectives.  The codebase also provides access to algorithms that have recently been added to the MPI standard, but are not yet provided in many system MPI installations, such as 
\begin{wrapfigure}[]{r}{0.2\textwidth}
  \vspace{-10pt}
  \includegraphics[width=0.15\textwidth]{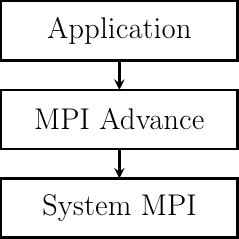}
  \vspace{-5pt}
  \caption{Dependencies}
  \vspace{-5pt}
  \label{fig:dep}
\end{wrapfigure}
partitioned communication and persistent collective operations.  MPI Advance also provides a library for GPU-Aware communication for heterogeneous architectures, providing the choice for utilizing GPUDirect when available, copying data to a single CPU, or multithreading data copied to the CPU to utilize all available cores. Finally, all MPI Advance libraries utilize the MPIX prefix in the user-facing APIs~\cite{skjellum1994extending,skjellum1994inter}.

MPI Advance sits between the application and system MPI installation, as exemplified in Figure~\ref{fig:dep}.  The library utilizes the MPIX-extensions, allowing for all applications to access MPI Advance optimizations without conflict.  MPI Advance then relies on the system MPI implementation for underlying message passing.

\vspace{-2mm}
\subsection{Collective Optimizations}
Collective algorithms have been optimized over multiple decades to minimize message count for small data sizes and bytes transported for larger messages.  In more recent years, architecture-aware algorithms further optimized many collectives by distinguishing between inter- and intra- node communication, reducing message count and size transported through the network.  Further optimizations, such as topology-aware collectives, gain additional performance by minimizing hop count.  MPI Advance provides an interface for a variety of these collective implementations in which the various algorithms are implemented on top of MPI, typically calling point-to-point communication within the underlying MPI implementation.

Currently, the collectives in this MPI Advance library select a default algorithm, but make all implementations publicly available so that users can select a non-default algorithm. Future work includes research into a more sophisticated selection process to help choose optimal collective algorithms for a given architecture and underlying MPI implementation. Listings 1 and 2 show how this library can be used to replace an existing all-to-allv operation inside an MPI application.

\vspace{-10pt}
\noindent\begin{minipage}{.42\textwidth}
\begin{lstlisting}[caption=Standard,frame=tlrb, language=C++]{Standard}
MPI_Alltoallv(sendbuf, sendcounts,
    sdispls, sendtype, recvbuf,
    recvcounts, rdispls, recvtype, 
    comm);
\end{lstlisting}
\end{minipage}\hfill
\begin{minipage}{.55\textwidth}
\begin{lstlisting}[caption=MPI Advance,frame=tlrb, language=C++]{MPI Advance}
MPIX_Comm* xcomm;
MPIX_Comm_init(&xcomm, comm);
MPIX_Alltoallv(sendbuf, sendcounts, sdispls, 
    sendtype, recvbuf, recvcounts, rdispls, 
    recvtype, xcomm);
\end{lstlisting}
\end{minipage}
\vspace{-14pt}

\subsection{Persistent Neighborhood Collective Optimizations}
Neighborhood collectives allow MPI to optimize sparse communication, in which each process communicates with a subset of other processes.  Persistent versions of neighborhood collectives were added to the MPI 4 standard, allowing for all setup costs to be incurred only once in the initialization method.  While standard implementations of neighbor collectives typically consist of simply wrapping point-to-point communication, the persistent API allows for optimizations of the underlying communication, such as locality-aware aggregation.  MPI Advance provides a library featuring locality-aware optimizations to persistent neighborhood collectives.

\vspace{-10pt}
\noindent\begin{minipage}{.42\textwidth}
\begin{lstlisting}[caption=Standard,frame=tlrb, language=C++]{Standard}
MPI_Comm comm;
MPI_Dist_graph_create_adjacent(
    comm, n_recvs, recv_procs, 
    recv_weights, n_sends, 
    send_procs, send_weights, 
    mpi_info, reorder, &comm);
MPI_Neighbor_alltoallv(sendbuf, 
    sendcounts, sdispls, sendtype, 
    recvbuf, recvcounts, rdispls, 
    recvtype, comm);
\end{lstlisting}
\end{minipage}\hfill
\begin{minipage}{.55\textwidth}
\begin{lstlisting}[caption=MPI Advance,frame=tlrb, language=C++]{MPI Advance}
MPIX_Comm xcomm;
MPIX_Request* xrequest;
MPIX_Dist_graph_create_adjacent(comm, 
    n_recvs, recv_procs, recv_weights, 
    n_sends, send_procs, send_weights, 
    mpi_info, reorder, &xcomm);
MPIX_Neighbor_alltoallv_init(sendbuf, 
    sendcounts, sdispls, sendtype, recvbuf, 
    recvcounts, rdispls, recvtype, xcomm, 
    info, &xrequest);
MPIX_Start(xrequest);
MPIX_Wait(xrequest, MPI_STATUS_IGNORE);
\end{lstlisting}
\end{minipage}

Persistent neighborhood collectives within this library can be used similarly to the collective operations previously described.  The neighborhood collective in Listing 3 only needs to be replaced with the persistent MPIX version in Listing 4.  Currently, there is an additional locality-aware extension to the neighborhood collective that requires additional data, namely unique indices for all values to be sent and received.  These unique values allow for the neighborhood collective to eliminate a single value from being sent between a set of nodes multiple times.

\vspace{-2mm}
\subsection{Partitioned Communication}
Partitioned collective communication \cite{MPI-4,finepoints}, is a new, channelized approach to point-to-point communication added in the MPI-4 standard.  Partitioned point-to-point communication establishes a single match between a sender and receiver at initialization, and supports the marking of portions of buffers (partitions), that can be transferred before the entire message is complete.  In this way, on the send side, partitioned sends allow so-called early-bird communication.  On the receive-side, partitions of the same or different sizes can be accessed as complete, prior to completion of the whole message.  In this paradigm, two-sided operations enable one-sided implementation internally, and support overlap of communication and computation in strong-progress implementations.  In situations where partitions are computed in parallel, such as in fork-join parallelism with OpenMP or CUDA kernels, partitioned communication helps hide the load imbalance of such computations and ensures that the computations and transfers have a good opportunity to overlap.

MPIPCL \cite{MPIPCL-JOURNAL} is a component of MPI Advance that supports the complete semantics of MPI-4 partitioned point-to-point operations with a reasonably performant interface.  The architecture of MPICL introduces a progress thread in order to move data asynchronously even when the underlying MPI implementation does not provide such a guarantee.  The semantics of initialization, initiation, and the Pready/Parrived partition operations are fully supported.  Simple partitioning strategies are also supported.  Significantly, MPICL can port on top of any MPI-3.0 compliant implementation that offers \texttt{MPI\_Thread\_multiple} modes of execution.  %
The audience for these APIs are early adopters of MPI+X modes of operation and implementers of partitioned operations in production MPI middleware. Additional extensions involving collective communication are being proposed for MPI-5 \cite{partcoll}. Similar to the point-to-point operations, a layered-library implementation of these collective APIs is being developed for MPIPCL~\cite{partcoll-impl} and will soon be available for public use.

\vspace{-2mm}
\subsection{Heterogeneous Architectures}
Emerging heterogeneous architectures often require applications to communicate data between GPU memories.  There are many different paths of data transfer, including using GPUDirect to move data directly from the GPU memory to the NIC, copying all data to a CPU before performing standard MPI communication between GPUs, copying portions of the data to each of the available CPU cores to distribute communication, or some combination of these protocols.  MPI Advance supports CUDA-aware and ROCM-aware communication, providing implementations for each of the various protocols mentioned above.  To utilize GPU-aware communication, this MPI Advance library must be compiled with either the CUDA or HIP flags.  Users can then call MPI operations, such as collectives and neighborhood collectives, passing either CPU or GPU memory.  While the libraries automatically select default implementations, all existing implementations are publicly available for the user to switch among them.

\section{Benefits and Results}~\label{section:results}
MPI Advance provides a lightweight interface that allows users to select specific implementations of methods within the MPI standard, create additional optimized implementations for existing methods, utilize new methods not yet within the standard or common installations of MPI, and perform MPI research without editing production MPI libraries.

Many of the optimizations within MPI Advance have been published in other papers, including locality-aware neighborhood collectives~\cite{collom2023optimizing}, locality-aware all-gather operations~\cite{BienzLocAwareBruck2022}, and optimized all-to-allv implementations~\cite{namugwanya2023collectiveoptimized}.  Furthermore, MPI Advance provides options for GPUDirect, copy-to-CPU, and copy-to-many-CPU approaches, as previously analyzed on Lassen and Summit, Power9 systems at LLNL and ORNL, respectively~\cite{HeteroModeling2021}.

MPIPCL has been similarly benchmarked. Comparisons against other MPI point-to-point operations~\cite{MPIPCL-FIRST} find that with only one partition, MPIPCL is no worse than base point-to-point operations. The second test was against an internal, RMA-based implementation of the partitioned point-to-point APIs~\cite{MPIPCL-JOURNAL}, finding that MPIPCL has similar performance, noting that MPIPCL may fall behind the internal implementation as it leverages more internal optimizations.

\section{Conclusions}~\label{section:conc}
Establishing community best practices, acceptance, and use of new and revised MPI features is crucial to maintaining its value as a key parallel programming model and continuing to deliver on its promise of performance-portability in the Exascale era. To that end, 
MPI Advance provides access to new standardized features in MPI on existing production systems that are not updated, and it provides access to experimental and future standard features long before they appear in the MPI standard or production implementations.  This set of capabilities helps speed up adoption and provides early feedback to the designers of new features before new editions of the MPI Standard are finalized.  

\begin{acks}
This work was performed with partial support from the National Science
Foundation under Grants Nos.~CCF-2151022, CCF-1918987, CCF-1562306,
CCF-1822191, CCF-1821431, OAC-1923980, OAC-1549812, OAC-1925603, OAC-2201497, and CCF-2151020 and the U.S. Department of Energy's National Nuclear Security Administration (NNSA) under the Predictive Science Academic Alliance Program (PSAAP-III), Award DE-NA0003966.  Any opinions, findings, and conclusions or recommendations expressed in this material are those of the authors and do not necessarily reflect the views of the NSF and the U.S. Department of Energy's NNSA.
\end{acks}

\bibliographystyle{ACM-Reference-Format}
\bibliography{refs}

\end{document}